\documentclass{aa}

\def\lsim{\lower.5ex\hbox{$\; \buildrel < \over \sim \;$}}
\def\gsim{\lower.5ex\hbox{$\; \buildrel > \over \sim \;$}}

\def\lsim{\lower.5ex\hbox{$\; \buildrel < \over \sim \;$}}
\def\gsim{\lower.5ex\hbox{$\; \buildrel > \over \sim \;$}}

\begin{document}
\input psbox.tex

\title{General relativistic effects of strong magnetic  fields on the
gravitational force: a driving engine for bursts of gamma rays in
SGRs?}

\titlerunning{General relativistic effects and SGRs}

\author{Manuel Malheiro\inst{1} \and Subharthi Ray\inst{1,2} \and
 Herman J. Mosquera Cuesta\inst{3} \and Jishnu Dey\inst{2,4}}

\authorrunning{Malheiro et al}

\offprints{Manuel Malheiro}

\institute{Instituto de Fisica, Universidade Federal Fluminense,
Niteroi 24210-340, RJ, Brazil\\
\and
Inter University Centre for Astronomy and Astrophysics,
Post Bag 4, Pune 411007, India\\
\and
Centro Brasileiro de Pesquisas F\'{\i}sicas, Laborat\'orio
de Cosmologia e F\'{\i}sica Experimental  de Altas Energias, Rio
de Janeiro 22290-180, RJ, Brazil\\
\and
UGC research professor, Department  of Physics, Maulana Azad
College, Kolkata 700013, India}

\date{}

\abstract{In general relativity all forms of energy contribute to
gravity and not only just ordinary matter as in Newtonian
Physics. This fact can be seen in the modified hydrostatic
equilibrium equation for relativistic stars pervaded by magnetic
(B) fields. It has an additional term coupled to the matter part
as well as an anisotropic term which is purely of magnetic origin.
That additional term coming from the pressure changed by the
radial component of the diagonal electromagnetic field tensor,
weakens the gravitational force when B is strong enough and can
even produce an {\it unexpected} change in the {\it attractive}
nature of the force by reversing its sign. In an extreme case,
this new general relativistic (GR) effect can even trigger  an
instability  in the star as a consequence of the {\it sudden}
reversal of the hydrostatic pressure gradient. We suggest here
that this GR effect may be the possible central engine driving
the transient giant outbursts observed in Soft Gamma-ray Repeaters
(SGRs). In small regions of the neutron star (NS), strong magnetic
condensation can take place. Beyond a critical limit, these
highly magnetised bubbles may explode releasing the trapped
energy as a burst of $\gamma$-rays of $\sim 10^{36-40}$ erg.
 }

\maketitle

\section{Introduction}

A  class  of  neutron  stars seem to have strong surface magnetic
fields of the order of $10^{15}$ Gauss. They are  referred  to  as
the Magnetars.  These  highly  magnetised  stars  have been
speculated to  be the progenitors of some {\it peculiar} gamma ray
bursts events, the so-called repeaters. Two kind of (candidates)
compact stars are believed to have very strong surface magnetic
field: the Soft Gamma-ray Repeaters (SGRs) and the Anomalous
X-ray Pulsars (AXPs: Mereghetti \& Stella 1995). SGRs are objects
that repeatedly emit bursts of gamma rays with energies of the
order of $10^{36-40}$ erg, in addition to persistent X-rays. The
SGRs are also a small enigmatic class of hard X-ray transient
sources (Marsden et al. 2001). AXPs show persistent X-ray
emission, modulated at a stable, slowly lengthening period.
Contrary  to the standard {\it binary X-ray pulsars}, they show
no evidence for a companion star.

Stars with strong surface magnetic fields have still stronger
field inside. Influence of strong magnetic fields on degenerate
nuclear matter has been studied  by Lai \& Shapiro (1991). They
employed the scalar virial theorem $ 2T + W +  3\Pi + \cal{M}=$0,
with $T$, $W$, $\Pi$ and $\cal{M}$, being the rotational kinetic
energy, the gravitational potential energy, the internal energy
and the magnetic  energy respectively. With this approach, they
found out that a surface field of $10^{13}$ Gauss can readily
imply that an  inner magnetic field to be as high as $10^{18}$
Gauss. Alternately, this field limit can easily be estimated if
we consider the magnetic energy approaching the gravitational
energy to produce a significant deviation from the normal balance
between pressure and gravity. This implies an extreme upper limit
for the average field B, with $(B^2/8\pi)(4\pi R^3/3)\sim GM^2/R$
so that $B_{lim} \sim 1.3 \, 10^8 (M/M_\odot)/(R/R_\odot)^2$
(Mestel, 1999). For a neutron star with $R/R_\odot \sim 10^{-5},$
$B_{lim} \sim 10^{18}$ Gauss. In this work we discuss the effect
on the gravitational force, of strong magnetic fields of the order
of the previously quoted limit.

Unlike its Newtonian counterpart, in the GR hydrostatic
equilibrium equation there appears a new term that depends not
only on the pressure of the gas  but also on the electromagnetic
field strength in the radial direction. When the  magnetic  field
is very strong and approaches  the critical limit,  it ``softens"
the gravitational force. In the extreme case, when the threshold
strength ($10^{18}$ Gauss) is reached, the sign of the radial
pressure gradient $\frac{dp_r}{dr} $ reverses, thus indicating
that the overall effect has become repulsive. An instability must
then appear, and a criterion for it can readily be set. The
appearance of a positive pressure gradient in the radial
direction, i.e., $\frac{dp_r}{dr} \ge 0 $, means that the NS is
unstable. This may be interpreted as if in the interior of the
magnetar the matter struggles to expand as the field reaches
(locally) such large values.

The electromagnetic field can be treated as a  fluid and there
exists room for the occurrence of magnetic field condensation in
a local volume in the NS interior. The  $B$ limit is called for
here to study highly magnetised matter bearing  in mind  a stable
stellar  configuration.  So, it is not surprising that in order
to reach the instability one needs to be nearly at this critical
field limit. The radial instability discussed here, and expected
to occur in the interior of a magnetar, is actually produced by
the {\it spatial anisotropy} of the magnetic stress-energy
tensor. The {\it anisotropic stress tensor} (i.e., the fact that
the radial component, $p_r $, differs from the angular
components, $p_\theta = p_\phi = p_t$ or, the transverse
component), introduces a {\it repulsive term}, purely of magnetic
origin, in the hydrostatic equilibrium equation. Besides this
term, the stellar {\it gravitational pull} is reduced because  it
depends directly on the radial component of the stress-energy
momentum tensor which becomes smaller in the presence of a very
strong magnetic field. Both effects conspire to make positive the
``effective" gradient of the radial pressure as to induce the
instability.

A  similar effect has been discussed recently by Ray et al.
(2003, 2004) with  regard  to  the  appearance of a  net electric
charge in a stellar structure. Therein, it is shown that there
exists a critical limit to the electric field, beyond which the
star is unstable. Here we make  use  of  this novel phenomenon,
the counterbalance of the gravitational pull excerpted by the
compact object in critically magnetised interiors of neutron
stars. The attentive reader must realize that not everywhere in
the interior of such stars the field is so high as $\sim 10^{18}$
Gauss, except only in small regions (the bubble) where it can be
raised by the effect of a strong magnetisation,  as discussed for
example by P\'erez Mart\'{\i}nez et al. (2003).  This instability
may trigger an abrupt release of energy similar to the giant
flares of soft gamma-ray outbursts seen in SGRs.

The mechanism for the emission of gamma rays from the SGRs is
still unknown. Typical energy of the emitted flares in the SGRs
is $10^{36}$ erg/s, with special exception of a few sources
emitting as high as $10^{44}$ erg/s. The magnetar model of
Wheeler et al. (2000) makes use of the high magnetic field
($10^{15}~ {\rm to}~ 10^{17}$ Gauss) of a rapidly rotating
neutron star as the possible source of cosmological Gamma-ray
bursts (GRBs) releasing energy of $\sim 10^{54}$ erg. As a
manifestation of the radial instability generated by the GR
effect discussed here, we also suggest a model to the emission
process of the SGRs, by the concentration of the magnetic field
in certain volume of the star, which we may call it as  a
``magnetic bubble". It is the fast expansion (explosion) of this
bubble, driven by the positive pressure gradient, which triggers
the quiescent state to the super outbursts in SGRs, and as such,
it may be a viable explanation of their central engine.

Magnetic field evolution in compact stars have been studied in
details by Thompson \& Duncan (1993, 1995, 1996 \& 2001) and by
Thompson, Lyutikov \& Kulkarni (2002). Their main idea for the
existence of high field in the neutron stars (magnetars) was, due
to some dynamo mechanism (alpha - Omega), based on the fact that
the neutron star convection is a transient phenomena and has
extremely high Reynolds number. They showed that most of the
magnetic energy of young pulsars reside in a convective cell of
diameter $\leq$ 1 km. Lyutikov \& Blandford (2002 \& 2003)
modeled the cosmological GRBs as explosions of electromagnetic
bubbles in a force free configuration, where the bubbles expand
non-relativistically inside the star, and then become highly
relativistic when they break from the surface.

Here also we take such an idea as a working hypothesis of the
formation of small volumes like bubbles, where the magnetic field
is very high and nearly close to some {\it critical field} limit.
The order of this critical magnetic field inside the magnetic
bubble is found to be the same as that of the pressure and the
mass of the inner sphere where the bubble is located. Also the
{\it not so frequent} giant flares found in SGR~0526$-$66, with
the famous March 5, 1979 event of energy release of $1.3 \times
10^{44}$ erg, and the SGR~1900$+$14, of the August 27, 1998 event
of energy release of $6.8 \times 10^{43}$ erg, can also be caused
by this relativistic effect of the strong magnetic field on the
gravitational force. We argue, conversely, that what the magnetic
field drives in fact is the reversal of the {\it gravitational
force} when a critical field strength is reached.

We organise this article in  the following way. In
Section~\ref{sec:formu} we present the formalism to describe the
stellar matter under strong magnetic fields in the light of
general relativistic formulation and present the modified
Tolman-Oppenheimer-Volkoff (TOV) equation. In the
Section~\ref{sec:instab}, we explain the instability and estimate
the critical value of B field as a function of the depth of the
star. In Section~\ref{sec:grb} we present the emission mechanism
of the bursts in SGRs and give the estimates of the released
energy and finally we draw our conclusions in
Section~\ref{sec:conc}.

\section{General relativistic effects of magnetic fields
on the structure equations}\label{sec:formu}

In this section we study the effect of the magnetic field in the
hydrostatic equilibrium equation. The magnetic field
energy-momentum tensor introduces an anisotropy in the fluid
pressure in the stellar interior. In our case we have an
anisotropy in the sense that the radial component of the pressure
$p_r $ differs from the angular components $p_\theta = p_\phi =
p_t$ (transverse pressure). We choose the magnetic field produced
by a magnetised surface to be a radial function which guarantees
that the pressure components are still a function of the radial
coordinate keeping the spherical symmetry.

We proceed with the standard form of the metric

\begin{equation}
ds^2=e^\nu   dt^2 -  e^\lambda dr^2  - r^2(d\theta^2  +
sin^2\theta d\phi^2). \label{eq:1}
\end{equation}

The Einstein-Maxwell stress tensor $T^\mu_\nu$ will have
contributions coming from the matter part and the magnetic
fields, and will take the form:

\begin{equation}
T^\mu_\nu  =   (p  +  \epsilon)u^\mu   u_\nu  -  p g^\mu_\nu
+ \frac{1}{4\pi} \left(-F^{\mu\alpha}F_{\alpha\nu} +
\frac{1}{4} g^\mu_\nu F_{\alpha\beta} F^{\alpha\beta}\right)
\label{eq:tmunu}
\end{equation}

where  p is  the pressure,  $\epsilon$ is  the energy  density
(=$\rho c^2$) and  $u$-s are the  4-velocity vectors. For the time
component, one  easily  sees  that  $u_t  = e^{\nu/2}$ and hence
$u^tu_t=1.$ Consequently, the other components (radial and
spherical) of the four vector are absent.

Maxwell's electromagnetic field equations can also be written as:

\begin{eqnarray}
F_{\mu \nu; \alpha}&=&0,
\label{eq:fmunu0}\\
{F^{\mu \nu}}_{;\nu}&=&4\pi j^{\mu},
\label{eq:fmunuj}
\end{eqnarray}
where $F_{\mu  \nu; \alpha}$ and $j^{\mu}$  are the covariant
derivative of the field strength tensor and the electric  current
4-vector, respectively. Eq. (\ref{eq:fmunu0}) gives the form of
the vector potential ($A_\mu$) as $F_{\mu \nu}=A_{\nu,
\mu}-A_{\mu, \nu}$.  So, the electric and magnetic fields in the
rest frame are defined as

\begin{eqnarray}
E_{\mu}&=&F_{\mu \nu} u^{\nu},
\label{eq:charge}\\
B_{\mu}&=&-\frac{1}{2}   \epsilon_{\mu  \nu   \alpha   \beta}  u^{\nu}
F^{\alpha \beta},
\label{eq:magnet}
\end{eqnarray}
$\epsilon_{\mu\nu\alpha\beta}$ being  the Levi-Civita
antisymmetric unit  tensor  with $\epsilon_{0123}=\sqrt{-g}$.
Now, considering the presence of a purely magnetic field, we can
write

\begin{equation}
E_{\mu}=F_{\mu \nu} u^{\nu}=0.
\label{eq:nocharge}
\end{equation}

This can also be thought of  as a highly conducting medium of the
star, where the term for the electrostatic charge will vanish.

The electromagnetic field strength is chosen in the form

\begin{eqnarray}
& & F_{\theta\varphi}=-F_{\varphi\theta}= B(r)\,r^2\,\sin\theta\,
,\label{mfield}
\end{eqnarray}
where $B(r)$ can be interpreted as the local magnetic field,  the
other components of $F_{\mu\nu}$ being zero.

Maxwell's equations (\ref{eq:fmunuj}) imply just one equation, the
$t$ component

\begin{equation}
\nabla_r F^{ti} = 4\pi J_m^t \; .
\end{equation}

Thus by using Eq.(\ref{mfield}) we have,

\begin{equation}
\left(r^2B\right)^\prime = 0\, , \label{maxdual2}
\end{equation}
where the prime stands for the derivative with respect
to the coordinate $r$. Such an equation can be integrated to give
$$B(r) = q_m/r^2,\quad r >0 \, , $$ $q_m$ being an integration
constant.


The nonzero components of the stress-energy tensor $T_\nu^\mu$ are

\begin{eqnarray}
\nonumber T_t^t = (\epsilon+\rho_{em}), \,~ T_r^r= -(p-\rho_{em})
\end{eqnarray}

and

\begin{eqnarray}
 T_\theta^\theta=T_\varphi^\varphi =-(p+\rho_{em})
\label{magtensor}
\end{eqnarray}
where $\rho_{em}=B^2/8\pi$ is the energy density in the
electromagnetic field. The relevant components of
Einstein-Maxwell field equations are

\begin{eqnarray}
& & \frac{e^{-\lambda}}{r^2}\left(r\lambda^\prime-1\right) +
  \frac{1}{r^2} = {8\pi } \left(\epsilon + \rho_{em}\right)
\, , \label{ein1}\\
& & \frac{e^{-\lambda}}{r^2}\left(r \nu^\prime+1\right) -
\frac{1}{r^2} = {8\pi }\left(p - \rho_{em}\right) \, .\label{ein2}
\end{eqnarray}

The first of the Einstein's  equations is used to determine the
metric coefficient $e^\lambda$, which is used to define the mass
and charge inside a sphere of radius $r$. Namely,

\begin{equation}
e^{-\lambda}=1-\frac{2m(r)}{r}+ \frac{q_m^2(r)}{r^2}\, ,
\label{lambda}
\end{equation}
where

\begin{eqnarray}
m(r)& =&\int^r_04\pi r^2\left({\epsilon}+ \rho_{em}\right)dr+
{q_m^2(r)\over 2r}\, .\label{mass}
\end{eqnarray}

The mass of the star is now due to the total contribution of  the
energy  density of the  matter and  the magnetic energy
($B^2/8\pi$) density. Such a definition is suitable since it
coincides with the mass of a localised object as observed from
infinity. So, for an observer at infinity, the mass is given by

\begin{eqnarray}
\nonumber
m_\infty &=&\int^R_04\pi r^2\left(\epsilon+
\rho_{em}\right)dr +
\int^\infty_R4\pi r^2\left(\epsilon+\rho_{em}\right)dr\\
&=&\int^R_04\pi r^2\left(\epsilon+ \rho_{em}\right)dr +
{q_m^2(R)\over 2R}  = m(R)\, , \label{minfty}
\end{eqnarray}
where R is the radius of the star.

 From Maxwell's equations we find $q_m(r)=B(r) r^2 - q_m(0)$. If
there are no singularities at $r=0$ the charge $q_m(0)$ is zero.
Therefore, it follows the well known relations for radial
magnetic field $B(r)= q_m/r^2$ (as above).

Exact solutions for anisotropic stars were obtained by Dev \&
Gleiser (2002;2003), where they considered anisotropy in the
pressure component for a spherically symmetric gravitationally
bound object. Taking a radial and transverse component of the
stress-energy tensor ($p_r$ and $p_t$ respectively), they arrived
at the analogous TOV equation for anisotropic stars.

\begin{eqnarray}
\frac{dp_r}{dr}=-\frac{(\rho+p_r)(M(r)+4\pi
r^3p_r)}{r^2(1-\frac{2M(r)}{r})} + \frac{2}{r}(p_t-p_r)\, .
\label{eq:aniso}
\end{eqnarray}

This equation explicitly shows the repulsive term coming from the
difference between the transverse and radial pressures (anisotropy
effect), with the {\it radial pressure} changing the
gravitational attraction. This is a purely relativistic effect
and it is of importance because of the appearance of a radial
magnetic field. To our knowledge, it has not been considered in
previous studies of magnetars.

Considering that $T^i_i = - p_i$ for $i=r,\theta~{\rm and}~ \phi$
and using in the equilibrium equation, Eq. (\ref{eq:aniso}), the
stress-energy momentum tensor from Eq. (\ref{magtensor}), and
replacing  the energy density $\rho $ by the time component
$T_t^t = (\epsilon+\rho_{em})$, the modified TOV for this case
will be:

\begin{eqnarray}
\frac{dp_r}{dr}=-\frac{\left(p+\epsilon\right)\left(M(r)+4\pi
r^3\left(p-\frac{B^2}{8\pi }\right)\right)}
{r^2\left(1-\frac{2M(r)}{r}\right)} + \frac{B^2}{2\pi r}
\label{eq:tovcomb}
\end{eqnarray}
where, the mass $M(r)$ is the mass of the inner sphere, below the
depth where the bubble is situated, and is integrated along the
entire length, from the center of the star to the bubble. Here
$M(r)$ is the mass for an observer in the star and is related to
the mass at infinity $m(r)$ from Eq. (\ref{mass}) by $M(r) = m(r)
-{q_m^2(r)/2r}$.

This form  for  the  modified hydrostatic equation equation is
essentially the same as has been recently done for the case of
presence of charge inside a star (Ray et al. 2003) if we
substitute the radial pressure gradient by  the total $dp/dr$
using the expression for $p_r = (p-\rho_{em})$ from Eq.
(\ref{magtensor}).


\section{A radial hydrostatic instability created by strong magnetic
fields}\label{sec:instab}

Let's find out the order of magnitude of the magnetic field so as
to produce an instability inside the star. Considering Eq.
(\ref{eq:tovcomb}), the negative sign of the radial pressure
gradient needs to be maintained for the entire radius of the star
for a stable configuration. The reversal of the force field
occurs only for a large value of the magnetic field. Now, the
first term on the right hand side of Eq. (\ref{eq:tovcomb}),
i.e., ($p+\epsilon$) is positive definite. So, the term which
primarily controls the `change of sign' for $dp_r/dr$ is
$(M(r)+4\pi r^3(p-B^2/8\pi))$. Also, the presence of the positive
term $B^2/(2\pi r)$ helps in the process of making the dp$_r$/dr
positive. So the stability criterion is

\begin{eqnarray}
\nonumber \frac{dp_r}{dr}\le 0
\end{eqnarray}
or,
\begin{eqnarray}
\frac{\left(p+\epsilon\right)\left(M(r)+4\pi r^3\left( p
-\frac{B^2}{8\pi}\right)\right)}
{r^2\left(1-\frac{2M(r)}{r}\right)} - \frac{B^2}{2\pi r} \ge 0 \,
. \label{eq:limit}
\end{eqnarray}

This can be written in terms of the limiting magnetic field as

\begin{equation}
B_{crit} \le \sqrt{2\pi(p+\epsilon)\frac{(M(r)+4\pi r^3
p)}{r\left(1-\frac{2M(r)}{r}+ \pi (p+\epsilon)r^2 \right)} }\, .
\label{eq:maxb}
\end{equation}

\begin{figure}[ht]
\vspace{-1.5cm} {\mbox{\psboxto(9cm;10cm){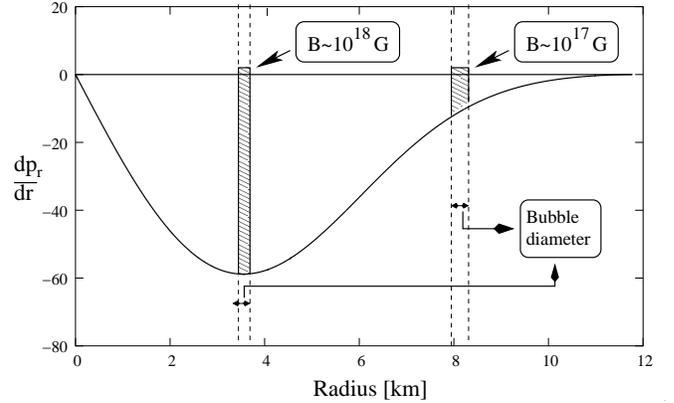}}} {\mbox{}}
\vspace{-2.5cm} \caption{The $dp_r/dr$ of the star as a function
of the radius of the star. The region of the space where it faces
a strong magnetic field, the $dp_r/dr$ becomes positive, and
produces a burst. The maximum strength of the field is necessary
where the depth of $dp_r/dr$ is the maximum.} \label{fig:dpdr}
\end{figure}

Magnetic field stronger than this critical field limit reverses
the sign of the radial pressure gradient, as shown in the Fig.
(\ref{fig:dpdr}).

What is the order of the critical magnetic field required to
equate both sides of the inequality (\ref{eq:maxb})? Taking
pressure ($p$) to be in $MeV/fm^3$, relation (\ref{eq:maxb}) shows
that the analog of B in that unit is $(MeV/fm^3)^{1/2}$. Hence,

\begin{eqnarray}
\nonumber
1~(MeV/fm^3)^{1/2}&\simeq& (1.6 \times 10^{33})^{1/2} c.g.s. ~units \\
&\simeq& 4 \times 10^{16} \rm Gauss. \label{eq:bconv}
\end{eqnarray}

Working out relation (\ref{eq:maxb}) we find that for the
characteristic pressure inside a neutrons star $p \sim 10
(MeV/fm^3) \sim 10^{35} dynes/cm^2$ the critical field limit is
$\sim 10^{2} (MeV/fm^3)^{1/2}$, which gives us, from relation
(\ref{eq:bconv}) a value of  $\sim 10^{18} Gauss.$ In this
extreme case, the magnetic stress $(B^2/8\pi)\sim 10^{35}
dynes/cm^2$ that is smaller but at the same order of the matter
pressure and can contribute to the stellar mass.

{\bf In normal matter, a non-potential magnetic field is held
inside a conductor in the form of stresses in the lattice. There
the critical stress is smaller than unity, and consequently $B^2$
would be much less than the pressure P. However, in a highly dense
and a complicated system like a neutron star where forces are
close to the strong interaction regime, the limit of the neutron
matter to support steady B field is still unknown and is expected
to be many orders of magnitude higher than that of normal matter.}

\begin{figure}[ht]
\vspace{-1.5cm} {\mbox{\psboxto(9cm;10cm){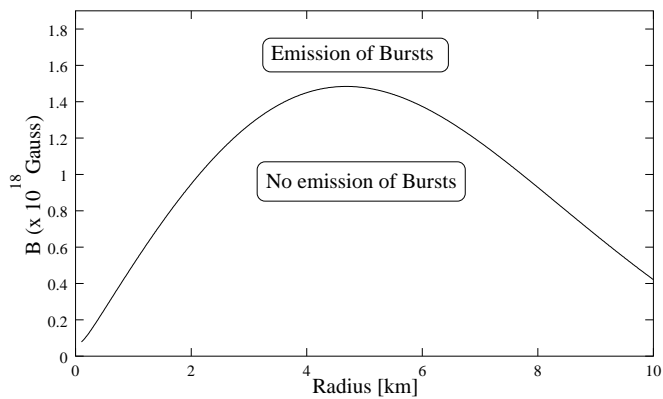}}} {\mbox{}}
\vspace{-2.5cm} \caption{Critical magnetic field (B) line defined
by the condition $dp_r/dr=0$ and given by Eq. (\ref{eq:maxb}).}
\label{fig:blim}
\end{figure}

The critical field as a function of the depth of the star is shown
in Fig. (\ref{fig:blim}). The choice of different EOSs will change
the number coefficients, but the overall order of magnitude will
remain the same. It is also shown in Fig. (\ref{fig:dpdr}) that
even for the maximum depth of the pressure gradient, the magnetic
field is calculated to be still of that order of magnitude. So,
the critical magnetic field depends essentially on the intrinsic
scales of the neutron star parameters, such as mass, pressure,
radius, etc.

{\bf It is worth mentioning at this point, that the strong B
field on $dp/dr$ (of Eq.(\ref{eq:tovcomb})) at $r$ indicates that
an entire shell of volume $4\pi r^2 dr$ should have a magnetic
field stronger than the critical field limit. However, alignment
of the magnetic moments of the nucleons, which is the possible
reason for the formation of the strong field, is not guaranteed
throughout the entire shell at the same time. So, the small
pockets inside the shell which gets magnetised, are referred here
as the magnetic bubbles, and the $dp/dr$ of Eq.(\ref{eq:tovcomb})
acts only on these small volume elements. The magnetic fields in
these bubbles are local.}


\section{Gamma-Ray emission process }\label{sec:grb}

One possible consequence of this general relativistic effect
coming from strong magnetic fields can be a mechanism for the
emission of gamma-ray flares in SGRs. Energy  emitted during each
burst process in a SGR is of the order of $10^{36}$  to $10^{40}$
erg with some special exceptions of about $10^{44}$ erg (the
March 5, 1979 event in SGR 0526$-$66 and the August 27, 1998
event in SGR 1900$+$14).

Fig. (\ref{fig:blim}) indicates the region below the curve as the
stable region, where there is no reversal of sign in the modified
TOV. Thus, this is the limit of the magnetic fields in the stars
having no burst. When the field limits exceeds the critical value
(as shown in Fig. (\ref{fig:blim})) of the magnetic field, then
we have flares of gamma-rays from the stars.

How much of the stellar  matter is needed  to produce a burst  of
the order of $10^{36}$ erg?  Typically, if we consider a small
bubble of 1 cm radius, then the volume of the bubble is $10^{-18}$
times than that of a neutron star of radius 10 km. For a bubble
of radius 1 cm, the total energy released will be

\begin{eqnarray}
{\cal E}= r^3 \frac{B^2}{6} \simeq(1)^3\times(10^{18})^2 erg
\simeq 10^{36} erg
\label{eq:estimate}
\end{eqnarray}
where we have used the scale of the order of $10^{18}$ Gauss as
obtained for the critical field strength.

So, to match with the energy released by the SGRs ($\sim 10^{36}$
erg/s), the bubble should have a volume of $\sim$ 1 cm$^3$. There
is a little variation in the scale due to the position of the
bubble inside the star. Another variation comes from the choice
of the EOS of the matter inside the star. These collected
variations however changes the numbers by maximally one order of
magnitude, but the general features will be restored. {\bf We
have considered a simple model where magnetic field condense in a
small sphere inside the neutron star.} So, this gives a very
general picture in our model of freezing of magnetic field inside
small droplets or bubbles, negligibly smaller than the volume of
the star, and emission of gamma-ray flares in the SGRs, from the
effect of relativistic equations. It can be equally applicable to
any and every neutron/strange star model having a strong magnetic
field.

\section{Conclusions}\label{sec:conc}

We have discussed a general relativistic effect, due to strong
magnetic fields, that can induce a radial instability inside
magnetars, with the subsequent release of gamma rays as an
outcome. We have estimated the strength of the magnetic field in
order to drive such an effect. The high magnetic field that is
necessary to change the chemical potential of the particles in
the nuclear matter of the neutron star has been calculated to be
more than 10$^{18}$ Gauss and the maximum mass of the entire
star, due to this high field, increases by a very small
percentage (e.g., see Table 2 \& 4 of Cardall et al., 2001). So,
for our model, a field of the order of $10^{17} {\rm to} 10^{18}$
Gauss inside a small bubble of radius 1 cm, which is more likely
to make a burst, the increase of chemical potential inside the
bubble due to very strong field, will leave very little effect on
the chemical potential in the rest of the matter of the star
where the field is comparatively weak. It is also worth
emphasising the fact that a freezing of the magnetic field in the
small volumes, is more likely due to alignment of magnetic
moments of the nucleons, rather than due to any convective
mechanism (e.g., Table 4 of Cardall et al., 2001), and can be
continuously produced. They do not need to form just during the
birth of the neutron star and survived for thousands of years.
Besides, a burst of this kind in the interior of a compact star
may produce a starquake as those already observed in the SGRs.
This relativistic effect due to the anisotropic nature of the
stress-energy momentum tensor inside a strongly magnetised NSs or
magnetars, as the progenitors for the central engine of the giant
flares in SGRs seems to having been overlooked in previous
studies.

\section*{Acknowledgments}
MM and HJMC thanks the nice scientific atmosphere at the
Workshop on Strong Magnetic Fields and Neutrons Stars at Havana,
Cuba where this work has began. MM thanks Vilson T. Zanchin for
valuable discussions and the CNPQ support. HJMC and SR acknowledges the
research support from FAPERJ (Brazil) and hospitality of the ICTP,
Trieste, Italy, where major part of the work was done. JD likes to
thank the UGC, Govt. of India.

\end{document}